\documentclass[aps,prb,preprint,groupedaddress,showpacs]{revtex4}
\usepackage{graphicx}
\usepackage{amssymb}
\begin{document}
\title{On the search for quantum criticality in a ferromagnetic system UNi$_{1-x}$Co$_x$Si$_2$}
\author{A.~P.~Pikul}
\author{D.~Kaczorowski}
\affiliation{Institute of Low Temperature and Structure Research, Polish Academy of Sciences, P Nr
1410, 50--590 Wroc{\l}aw 2, Poland}
\date{\today}

\begin{abstract}
Polycrystalline samples of the isostructural alloys UNi$_{1-x}$Co$_x$Si$_2$ ($0 \leqslant x
\leqslant 1$) were studied by means of x-ray powder diffraction, magnetization, electrical
resistivity and specific heat measurements, at temperatures down to 2~K and in magnetic fields up
to 5~T. The experimental data revealed an evolution from strongly anisotropic ferromagnetism with
pronounced Kondo effect, observed for the alloys with $x < 0.98$ and being gradually suppressed
with rising Co-content, to spin-glass-like states with dominant spin fluctuations, seen for the
sample with $x=0.98$. Extrapolation of the value of $T_{\rm C}(x)$ yields a critical concentration
$x_{\rm c} = 1$, at which the magnetic ordering entirely disappears. This finding is in line with
preliminary data collected for stoichiometric UCoSi$_2$.
\end{abstract}
 \maketitle

\section{Introduction}

Since a few years the unusual physical behaviors of $f$-electron systems being at the verge of
magnetic ordering have been at the very forefront of modern condensed matter physics
[\onlinecite{stewart1,stewart2,coleman1,loehneysen,gegenwart2}]. The so-called quantum phase
transition (QPT) occurs at the absolute zero temperature and is driven by quantum fluctuations,
instead of thermal fluctuations associated with phase transitions at finite temperatures. If such a
transition has a continuous (2nd order) character one speaks about a quantum critical point (QCP).

Most of the hitherto performed experimental studies on the quantum criticality (independently of
the theoretical scenarios considered) were devoted to numerous non-stoichiometric alloys and
stoichiometric compounds based on cerium, ytterbium or uranium, which exhibit antiferromagnetic
correlations [\onlinecite{stewart1,stewart2}]. Ferromagnetic quantum phase transitions are much
less known both from the theoretical and experimental point of views
[\onlinecite{stewart1,stewart2,kirkpatrick}]. Therefore, it is particularly tempting to investigate
systems in which ferromagnetism can be tuned down to the absolute zero temperature by external
parameters, like pressure, magnetic field or/and composition. Amidst very rare examples of alloys,
in which ferromagnetism gets suppressed (or induced) by chemical doping, one can mention i.a.
U$_x$Th$_{1-x}$Cu$_2$Si$_2$ [\onlinecite{lenkewitz}], Ni$_{x}$Pd$_{1-x}$ [\onlinecite{nicklas}],
CePd$_{1-x}$Ni$_{x}$ [\onlinecite{kappler}],  CePd$_{1-x}$Rh$_{x}$
[\onlinecite{sereni,pikul2,westerkamp}], URu$_{2-x}$Re$_x$Si$_2$ [\onlinecite{bauer}],
URh$_{1-x}$Ru$_x$Ge [\onlinecite{huy}] and UCoGe$_{1-x}$Si$_{x}$ [\onlinecite{denijs}]. The number
of stoichiometric compounds for which the presence  of pressure-induced ferromagnetic QPT has been
demonstrated is extremely small, with the most prominent examples being UGe$_2$
[\onlinecite{saxena}], URhGe [\onlinecite{aoki}], UCoGe [\onlinecite{huy2}]. Remarkably, up to date
the only stoichiometric ferromagnet claimed to exhibit a second-order quantum phase transition
seems CePt [\onlinecite{larrea}].

The ternary uranium silicides UTSi$_2$ (T = Fe, Co, Ni), crystallizing with the orthorhombic
CeNiSi$_2$-type crystal structure, were reported to span a variety of magnetic properties driven by
the hybridization between U $5f$- and T $3d$-electronic states. Whereas UNiSi$_2$ is a
ferromagnetically ordered ($T_{\rm C}$~=~95~K) Kondo lattice with relatively well localized
$5f$-electrons [\onlinecite{kaczorowski2,taniguchi,das}], UCoSi$_2$ behaves as a spin fluctuation
system, and UFe$_{1-y}$Si$_2$ shows features of a weakly temperature dependent Pauli paramagnet
[\onlinecite{kaczorowski2}]. The previous findings motivated us to undertake a detailed study on
the solid solution UNi$_{1-x}$Co$_x$Si$_2$ ($0 \leqslant x \leqslant 1$), with the main focus on
the alloys being close to a ferromagnetic instability, expected to occur for a~certain Co-content
$x_{\rm c}$. Our first attempt was to check for a possible non-Fermi-liquid character of the dc
magnetic susceptibility, electrical resistivity and heat capacity of the specimens having nearly
critical composition. The aim of this paper is to show that the ferromagnetic behavior in
UNi$_{1-x}$Co$_x$Si$_2$ is observed in a nearly complete solution range and that the critical
concentration $x_{\rm c}$ is fairly close to 1. In other words, our experimental data point at
possible ferromagnetic quantum critical behavior in stoichiometric UCoSi$_2$.

\section{Experimental details}
Polycrystalline samples of the solid solutions UNi$_{1-x}$Co$_x$Si$_2$ ($0 \leqslant x \leqslant
1$) were synthesized by conventional arc melting the nominal amounts of the constituents under
protective atmosphere of an argon glove box. The pellets were subsequently wrapped in a tantalum
foil, sealed in evacuated silica tubes, and annealed at 800$^{\circ}$C for 2 weeks. Quality of the
products was verified by means of x-ray powder diffraction measurements (X'pert Pro PANalytical
diffractometer with Cu K$\alpha$ radiation; $\lambda$~=~1.54056~\AA). Magnetic properties were
studied at temperatures ranging from 1.7 up to room temperature and in applied magnetic fields up
to 5~T, using a Quantum Design SQUID (superconducting quantum interference device) magnetometer and
a Cryogenics AC susceptometer. The electrical resistivity was measured down to 4.2~K in zero
magnetic field using a standard DC four-point probe method implemented in a home-made setup, on
bar-shaped specimens with spot-welded electrical contacts. Heat capacity studies were carried out
at temperatures 2~K~--~300 K and in applied magnetic fields up to 9 T employing a thermal
relaxation technique implemented in a Quantum Design PPMS (Physical Property Measurement System).

\section{Results and discussion}

\subsection{Crystal structure}

\begin{figure}
\includegraphics[width=7cm]{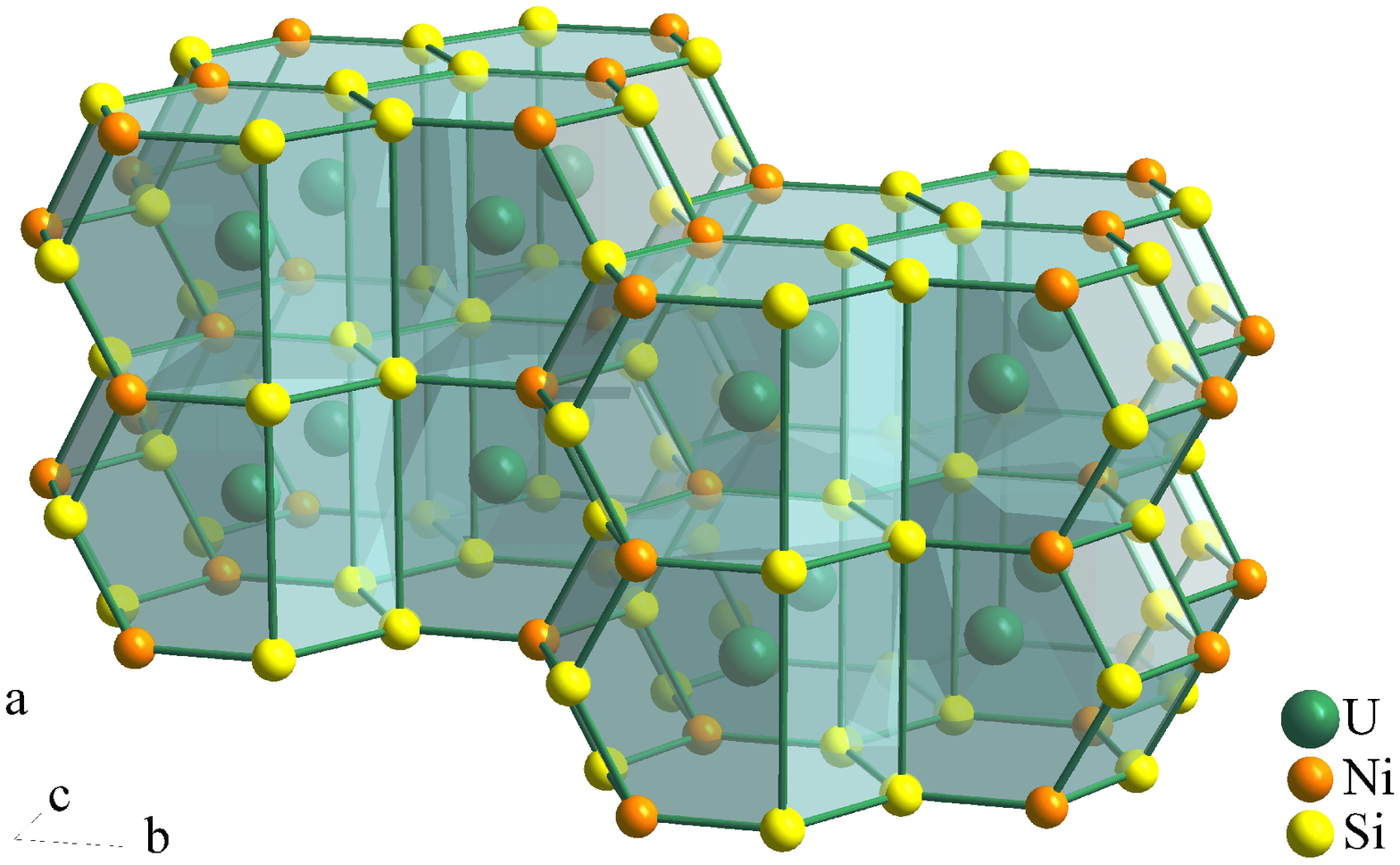}%
\\ \vspace{1em}
\includegraphics[width=7cm]{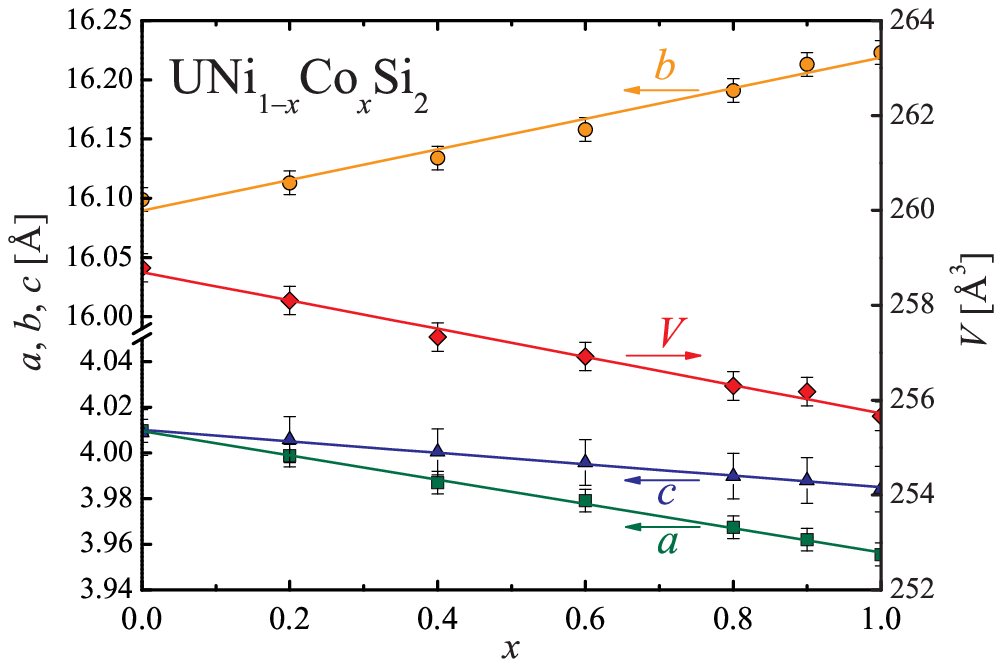}%
\caption{\label{fig_struct} (Color online) (a) Crystal structure of UNiSi$_2$. (b) Lattice parameters
and unit cell volume of the solid solutions UNi$_{1-x}$Co$_x$Si$_2$ as a function of the Co content $x$.}
\end{figure}

Analysis of the experimental x-ray powder diffraction patterns obtained for the alloys
UNi$_{1-x}$Co$_x$Si$_2$ (not shown here) revealed that all the specimens used in the present study were
nearly single phases with very small amount (less than 5\%) of same unidentified impurity phase. Rietveld
refinements confirmed that the entire system crystallizes with the orthorhombic ($Cmmm$) CeNiSi$_2$-type
structure [Fig.~\ref{fig_struct}(a)]. The lattice parameters derived for UNiSi$_2$ ($a$~=~4.006~\AA, $b$~=~16.070~\AA
~and $c$~=~4.002~\AA) are in good agreement with the literature data [\onlinecite{kaczorowski2}]. As can be
inferred from Fig.~\ref{fig_struct}(b), the isostructural, partial substitution of Ni atoms by
about 0.8\% larger Co atoms expands linearly the unit cell along the $b$ axis (in total by $+0.8$\%)
yet reduces it along the $a$ and $c$ axes (in total by $-1.4$\% and $-0.6$\%, respectively), leading
eventually to the contraction in the unit cell volume of the system by about $-1.2$\%. The lattice
parameters of the terminal compound UCoSi$_2$ were found to be equal to $a$~=~3.956~\AA,
$b$~=~16.223~\AA ~and $c$~=~3.984~\AA, being close to those reported previously
[\onlinecite{kaczorowski2}].

\subsection{Magnetic properties}

\begin{figure}
\includegraphics[width=7cm]{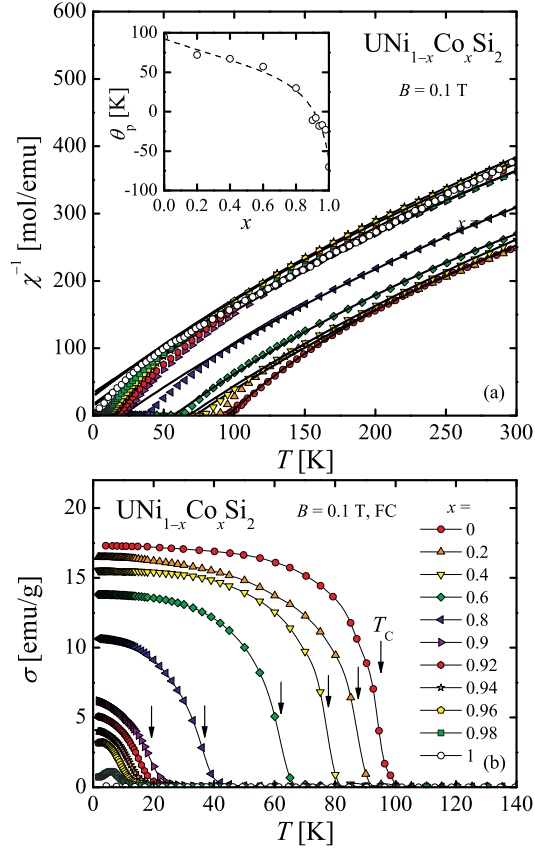}%
\caption{\label{fig_suscept_all} (Color online) (a) Temperature dependencies of the inverse molar magnetic
susceptibility of the alloys UNi$_{1-x}$Co$_{x}$Si$_2$ measured in magnetic field $B$~=~0.1~T. Solid lines are fits
of the modified Curie--Weiss law (Eq.~(\ref{eq_mCW})) to the experimental data. The inset displays the Weiss temperature
$\theta_{\rm p}$ as a function of $x$; the dashed line serves as a guide for the eye.
(b) Temperature variations of the magnetization of UNi$_{1-x}$Co$_{x}$Si$_2$
measured in $B$~=~0.1~T in a field-cooling (FC) regime. Thin solid
lines serve as guides for the eye; the arrows mark the ordering temperatures $T_{\rm C}$.}
\end{figure}

Figure \ref{fig_suscept_all}(a) presents the inverse molar magnetic susceptibility $\chi^{-1}$ of
the alloys UNi$_{1-x}$Co$_x$Si$_2$ ($0 \leqslant x \leqslant 1$) as a function of temperature $T$.
As seen, above about 150~K all the experimental curves exhibit quasi-linear behavior that can be
described by the modified Curie--Weiss law:
\begin{equation}
\chi (T) = \frac{1}{8} \frac{\mu_{\rm eff}^2}{T-\theta_{\rm p}} + \chi_0,
\label{eq_mCW}
\end{equation}
where $\mu_{\rm eff}$ is the effective magnetic moment, $\theta_{\rm p}$ is the paramagnetic Weiss
temperature, and $\chi_0$ is a temperature independent term accounting for Pauli paramagnetism of
the conduction electrons and diamagnetism of the core electrons. Least squares fits of
Eq.~(\ref{eq_mCW}) to the experimental data [see the solid lines in Fig.~\ref{fig_suscept_all}(a)]
yielded for each UNi$_{1-x}$Co$_x$Si$_2$ compound a value of $\mu_{\rm eff} \sim 2.3\,\mu_{\rm B}$
and $\chi_0 \sim 8\times 10^{-4}$ emu/mol. In turn, $\theta_{\rm p}$ was found to decrease
monotonically from 95~K in UNiSi$_2$ to $-70$~K in UCoSi$_2$ [see the inset to
Fig.~\ref{fig_suscept_all}(a)]. The experimental value of $\mu_{\rm eff}$ is moderately reduced in
comparison to the theoretical one, calculated for free U$^{3+}$ (3.62~$\mu_{\rm B}$) or U$^{4+}$
(3.58~$\mu_{\rm B}$) ions, and $\chi_0$ is somewhat larger than usually observed in conventional
metals. It suggests that the uranium $5f$ electrons in UNi$_{1-x}$Co$_x$Si$_2$ are partly
delocalized, as commonly observed in U-based intermetallics. Because $\mu_{\rm eff}$ and $\chi_0$
hardly changes upon increasing $x$ one can conclude, that the degree of delocalization of the $5f$
electrons in this system remains unaffected by the Ni/Co substitution. In turn, the decreasing
positive $\theta_{\rm p}$ can be ascribed to weakening of an inter-site ferromagnetic coupling
between the uranium magnetic moments. For large $x$ it becomes obscured by a negative contribution
most likely due to strong electronic correlations or an inter-site antiferromagnetic coupling.

Figure \ref{fig_suscept_all}(b) displays the temperature dependencies of the magnetization $\sigma$
of UNi$_{1-x}$Co$_x$Si$_2$. The characteristic Brillouin-shaped anomalies on $\sigma(T)$ indicate
the ferromagnetic ordering in the alloys with $x \leqslant 0.96$. Upon increasing the Co content,
the Curie temperature $T_{\rm C}$ decreases from 95~K in UNiSi$_2$ (in agreement with the
literature data [\onlinecite{kaczorowski2}]) down to 8.6~K in UNi$_{0.04}$Co$_{0.96}$Si$_2$.
Simultaneously, the magnitude of the magnetization notably decreases. Eventually, for $x > 0.96$,
the ferromagnetic  anomaly evolves into a broad maximum of unclear origin, while in pure UCoSi$_2$
no anomaly in $\sigma(T)$ is observed down to 1.7~K.

\begin{figure}
\includegraphics[width=7cm]{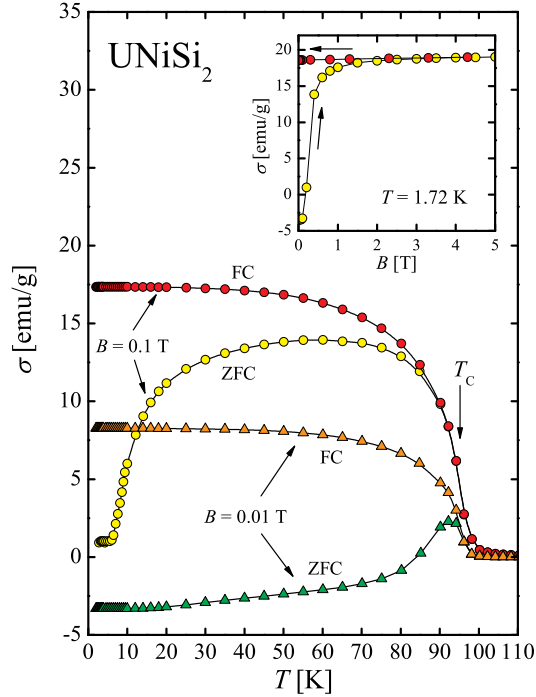}%
\caption{\label{fig_suscept_0} (Color online) Temperature variations of the magnetization $\sigma$ of UNiSi$_2$
measured in two different magnetic fields $B$ in the zero-field-cooling (ZFC) and the field-cooling
(FC) regime. The solid lines serve as guides for the eye; $T_{\rm C}$ marks the Curie temperature.
The inset: $\sigma$ measured at constant temperature $T$~=~1.72~K as a function of increasing (open symbols) and
decreasing (closed symbols) magnetic field $B$.}
\end{figure}

\begin{figure}
\includegraphics[width=7cm]{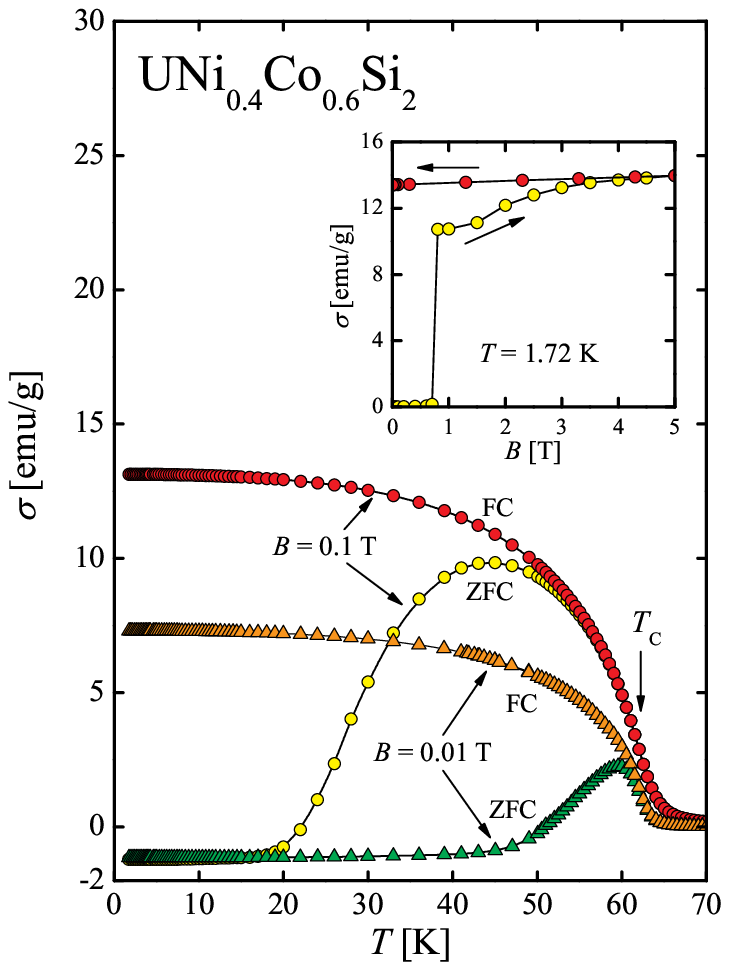}%
\caption{\label{fig_suscept_06} (Color online) Temperature variations of the magnetization $\sigma$ of UNi$_{0.4}$Co$_{0.6}$Si$_2$
measured in two different magnetic fields $B$ in the zero-field-cooling (ZFC) and the field-cooling (FC) regime.
The solid lines serve as guides for the eye;  $T_{\rm C}$ marks the Curie temperature. The inset:
$\sigma$ measured at constant temperature $T$~=~1.72~K as a function of increasing (open symbols) and
decreasing (closed symbols) magnetic field $B$.}
\end{figure}

The results of detailed investigations of the magnetic properties of two representative
compositions from the range $x \leqslant 0.96$, namely UNiSi$_2$ and UNi$_{0.4}$Co$_{0.6}$Si$_2$,
are displayed in Figs.~\ref{fig_suscept_0} and \ref{fig_suscept_06}, respectively. For both
compounds $\sigma (T)$ measured in zero-field-cooled and field-cooled regimes shows a pronounced
bifurcation below the Curie temperature, characteristic of strongly anisotropic ferromagnets with
pronounced domain effects. The negative values of the ZFC magnetization result most likely from the
presence of remnant magnetic field, hardly avoidable in experiments performed using standard
superconducting magnets. The overall shape of the isothermal field dependence of the magnetization
measured deeply  in the ordered state [see the insets to Figs.~\ref{fig_suscept_0} and
\ref{fig_suscept_06}] provides another strong evidence for the ferromagnetic ground state in
UNiSi$_2$ and UNi$_{0.4}$Co$_{0.6}$Si$_2$. In particular, in low magnetic fields the magnetization
rapidly increases with $B$ and saturates at high fields, reaching at $B$ = 5~T a value
corresponding to about 1.2~$\mu_{\rm B}$ (UNiSi$_2$) and 0.9~$\mu_{\rm B}$
(UNi$_{0.4}$Co$_{0.6}$Si$_2$). Remarkably, the two $\sigma (B)$ curves exhibit strongly hysteretic
behavior and very high remanence of about 19 (for $x = 0.0$) and 13.5 emu/g (for $x = 0.6$), both
features being characteristic of hard ferromagnets. The magnetic properties of the other alloys
UNi$_{1-x}$Co$_x$Si$_2$ from the range $x \leqslant 0.96$ are qualitatively very similar to those
of UNiSi$_2$ and UNi$_{0.4}$Co$_{0.6}$Si$_2$, yet the hysteresis and the saturated moment both
gradually diminish with increasing $x$.

\begin{figure}
\includegraphics[width=7cm]{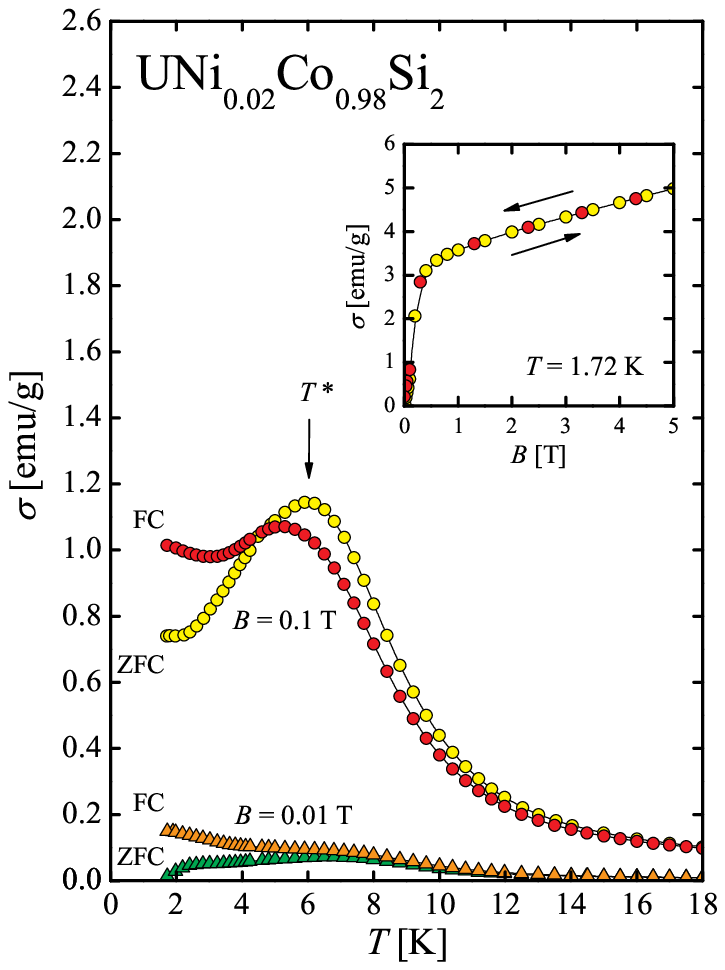}%
\caption{\label{fig_suscept_098} (Color online) Temperature variations of the magnetization $\sigma$
of UNi$_{0.02}$Co$_{0.98}$Si$_2$ measured in two different magnetic fields $B$ in the zero-field-cooling (ZFC)
and the field-cooling (FC) regime. The solid lines serve as guides for the eye and the arrows mark
the phase transition temperatures. The inset: $\sigma$ measured at constant temperature $T$~=~1.72~K
as a function of increasing (open symbols) and decreasing (closed symbols) magnetic field $B$.}
\end{figure}

Figure~\ref{fig_suscept_098} presents the magnetic properties of UNi$_{0.02}$Co$_{0.98}$Si$_2$.
Apparently, the overall shape of the FC $\sigma (T)$ curve measured for this composition is much
different from that observed for the ferromagnetically ordered alloys (i.e. with $x \leqslant
0.96$). In particular, the Brillouin-like curvature is not seen, and instead the $\sigma (T)$ curve
forms a broad maximum at about 6~K. Moreover, the magnitude of the magnetization of
UNi$_{0.02}$Co$_{0.98}$Si$_2$ is almost 20 times smaller than the value measured for the parent
compound UNiSi$_2$ [Fig.~\ref{fig_suscept_0}], the hysteresis in $\sigma (B)$ [see the inset to
Fig.~\ref{fig_suscept_098}] is hardly visible, and $\sigma$ does not saturate in high fields. All
these features indicate a change in the character of the magnetic ordering from the long-range
ferromagnetic one for $x \leqslant 0.96$ to a spin-glass-like state for $x = 0.98$, probably
governed by an interplay of competing ferromagnetic and antiferromagnetic interactions.

In agrement with the previous report~[\onlinecite{kaczorowski2}], the terminal alloy UCoSi$_2$ is
a Curie-Weiss paramagnet down to the lowest temperature studied (cf. Fig.~\ref{fig_suscept_all}).
Accordingly, its magnetization does not exhibit any hysteresis effect neither in the field nor the
temperature dependence (not shown here).

\subsection{Electrical resistivity}

\begin{figure}
\includegraphics[width=7cm]{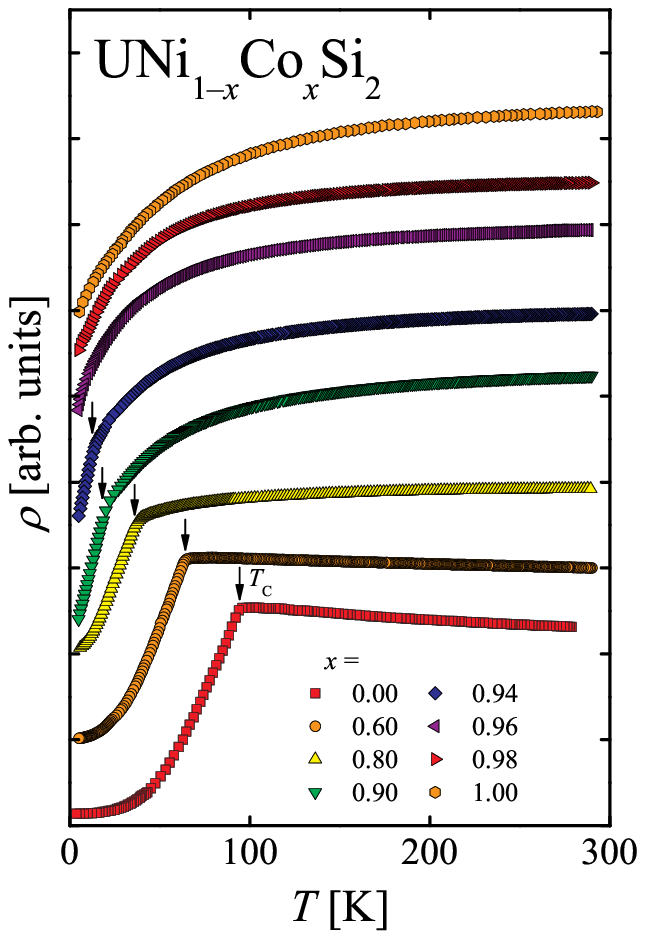}%
\caption{\label{fig_resist_all} (Color online) Temperature dependencies of the electrical resistivity $\rho$
of the alloys UNi$_{1-x}$Co$_x$Si$_2$ with selected Co-content; the curves are shifted upwards for clarity.
The arrows mark the ordering temperatures $T_{\rm C}$ and the solid lines emphasize logarithmic temperature dependence of $\rho$.}
\end{figure}

Temperature dependencies of the electrical resistivity $\rho$ of the UNi$_{1-x}$Co$_x$Si$_2$ alloys
are plotted in Fig.~\ref{fig_resist_all}. In the paramagnetic region, the electrical resistivity of
the alloys with $x \leqslant 0.60$ increases with decreasing temperature. Upon further substitution
of Ni atoms by Co atoms, the $\rho (T)$ curve changes its slope into a positive one in the entire temperature range studied.

\begin{figure}
\includegraphics[width=7cm]{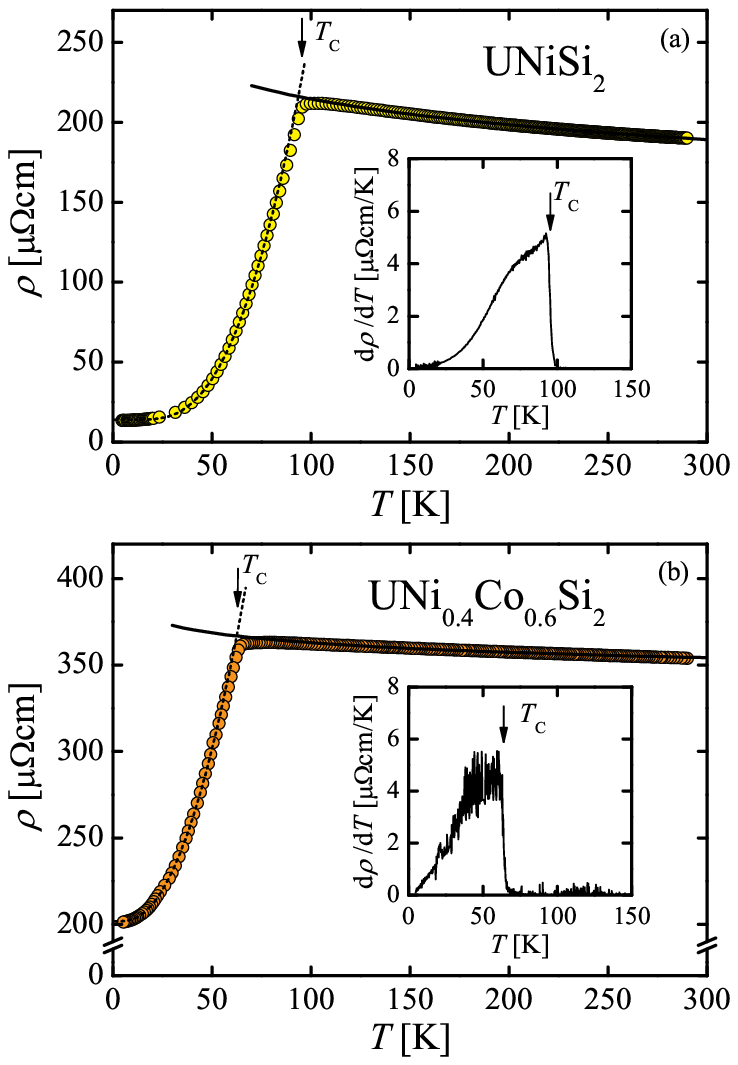}%
\caption{\label{fig_resist_1_0.6} (Color online) Temperature dependence of the electrical resistivity $\rho$ of (a) UNiSi$_2$ and (b) UNi$_{0.4}$Co$_{0.6}$Si$_2$. The arrows mark the ordering temperatures $T_{\rm C}$, the solid
and the dashed line show fits of Eqs.~(\ref{eq_res_log_kondo}) and (\ref{eq_res_magnons}), respectively, to the experimental data.
The insets show temperature derivatives of the resistivity, ${\rm d}\rho (T)/{\rm d}T$; the arrows mark the Curie
temperatures $T_{\rm C}$.}
\end{figure}

At low temperatures, pronounced drops of the resistivity manifest the ferromagnetic phase
transitions in the alloys with $x \leqslant 0.94$. Positions of the anomalies in $\rho (T)$,
defined as inflection points in ${\rm d}\rho(T)/{\rm d}T$ (cf. Fig.~\ref{fig_resist_1_0.6}),
correspond to the Curie temperatures estimated from the magnetization data [cf.
Fig.~\ref{fig_suscept_all}(b)]. For the Co-rich alloys UNi$_{0.04}$Co$_{0.96}$Si$_2$ ($T_{\rm
C}$~=~8.6~K) and UNi$_{0.02}$Co$_{0.98}$Si$_2$ ($T^{\ast}$~=~6~K), the phase transitions are hardly
visible in the resistivity data.

The increase of the resistivity with decreasing $T$, that is observed for $x \leqslant 0.60$ in the
paramagnetic region, may result from the spin-flip Kondo scattering the conduction electrons on magnetic
moments of the uranium atoms. Indeed, the experimental data can be described above $T_{\rm C}$ by
the formula [\onlinecite{kaczorowski2}]:
\begin{equation}
\rho (T) = \left( \rho_0 + \rho_0^{\infty} \right) - c_{\rm K} \log{T},
\label{eq_res_log_kondo}
\end{equation}
where the first term stands for temperature independent scattering the conduction electrons on
static defects and disordered magnetic moments, and the second one describes the Kondo effect.
Solid lines in Fig.~\ref{fig_resist_1_0.6} display the results of the least-square fits for
UNiSi$_2$ [$\left( \rho_0 + \rho_0^{\infty} \right)$~=~319.6(1)~$\mu \Omega$cm, $c_{\rm
K}$~=~52.6(1)~$\mu \Omega$cm] and UNi$_{0.4}$Co$_{0.6}$Si$_2$ [$\left( \rho_0 + \rho_0^{\infty}
\right)$~=~396.2(1)~$\mu \Omega$cm, $c_{\rm K}$~=~16.9(1)~$\mu \Omega$cm]. The values of the
fitting parameters for UNiSi$_2$ are of the same order as hose given in
Ref.~\onlinecite{kaczorowski2}.  The observed decrease in the value of $c_{\rm K}$ on going from
UNiSi$_2$ to UNi$_{0.4}$Co$_{0.6}$Si$_2$ manifests suppression of the Kondo effect in the system
with increasing $x$, in line with the observed evolution of $\rho(T)$ towards strongly bent
metallic-like dependence found for the Co-rich samples.

In the ordered region, the resistivity of both alloys is dominated by the contributions due to
scattering the conduction electrons on ferromagnetic spin waves [\onlinecite{kaczorowski2}]:
\begin{equation}
\rho (T) = \rho_0 + A T^2 \exp{\left( -\frac{\Delta}{T} \right)},
\label{eq_res_magnons}
\end{equation}
where $\rho_0$ is the residual resistivity, $\Delta$ is a gap in the spin-waves spectrum, and
$c_{\rm m}$ is a coefficient of proportionality. Fits of Eq.~(\ref{eq_res_magnons}) to the
experimental data below about $0.8~T_{\rm C}$ yielded the values $\rho_0$~=~13.9~$\mu \Omega$cm,
$c_{\rm m}$~=~0.06(1)~$\mu \Omega$cm/K$^2$ and $\Delta$~=~87(1)~K for UNiSi$_2$, and
$\rho_0$~=~201.8~$\mu \Omega$cm, $c_{\rm m}$~=~0.05(1)~$\mu \Omega$cm/K$^2$ and $\Delta$~=~15(1)~K
for UNi$_{0.4}$Co$_{0.6}$Si$_2$ (see the dashed lines in Fig.~\ref{fig_resist_1_0.6}). The values
obtained for UNiSi$_2$ somewhat differ from those reported in Ref.~\onlinecite{kaczorowski2}, most
probably due to distinctly different ranges of the data analysis.

\begin{figure}
\includegraphics[width=7cm]{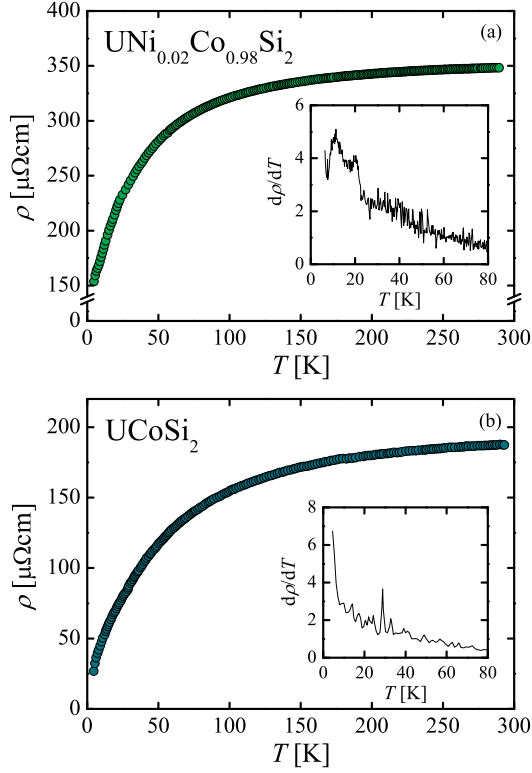}%
\caption{\label{fig_resist_0.98_0.00} (Color online) Temperature dependence of the electrical resistivity
$\rho$ of (a) UNi$_{0.02}$Co$_{0.98}$Si$_2$ and (b) UCoSi$_2$. The insets show temperature derivatives ${\rm d}\rho (T)/{\rm d}T$.}
\end{figure}

Figure \ref{fig_resist_0.98_0.00} presents the temperature dependencies of the electrical
resistivity of the Co-rich sample UNi$_{0.02}$Co$_{0.98}$Si$_2$ and the terminal compound
UCoSi$_2$. As seen, the overall shapes of these two $\rho(T)$ curves are reminiscent of metallic
systems with  strong spin fluctuations, like e.g. Np, Pu, or UAl$_2$. Both $\rho(T)$ and the
derivative ${\rm d}\rho (T)/{\rm d}T$ calculated for UCoSi$_2$ are featureless down to the lowest
temperatures studied. Also for UNi$_{0.02}$Co$_{0.98}$Si$_2$ no anomaly in the electrical transport
is seen, which could be associated with the feature at $T^{\ast}$ revealed in the magnetic
susceptibility data (broad and blurred extremum observed in ${\rm d}\rho (T)/{\rm d}T$ is likely
due to experimental noise).

\subsection{Specific heat}

\begin{figure}
\includegraphics[width=7cm]{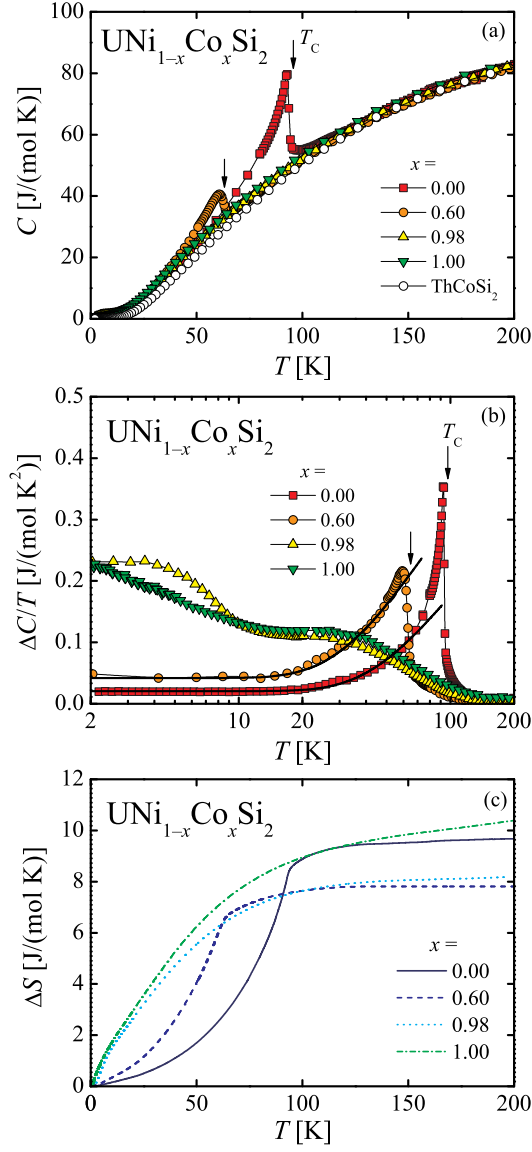}%
\caption{\label{fig_heat_all} (Color online) (a) Temperature dependencies of the specific heat $C$ of selected
UNi$_{1-x}$Co$_x$Si$_2$ alloys and ThCoSi$_2$. The arrows mark the ordering temperatures $T_{\rm C}$.
(b) Magnetic contribution $\Delta C$ divided by $T$ in logarithmic scale. Solid lines are fits of
Eq.~(\ref{eq_hc_magnons}) to the experimental data. (c) Temperature dependencies of the excess magnetic
entropy $\Delta S$ in selected UNi$_{1-x}$Co$_x$Si$_2$ alloys. The arrows mark the ordering temperatures
$T_{\rm C}$.}
\end{figure}

Figure~\ref{fig_heat_all}(a) presents the temperature variations of the specific heat of selected
UNi$_{1-x}$Co$_x$Si$_2$ alloys and their isostructural phonon counterpart ThCoSi$_2$. Above about
110~K, the heat capacity data obtained for each alloy overlap the $C(T)$ curve of the latter
compound, indicating that in this temperature range the possible Schottky contribution due to
crystal electric field is negligibly small in comparison to the total specific heat measured.
Presence of a distinct $\lambda$-shaped peak in $C(T)$ at $T_{\rm C}$ confirms bulk character of
the long-range ferromagnetic ordering in UNiSi$_2$. For UNi$_{0.40}$Co$_{0.60}$Si$_2$ the magnetic
anomaly is still clearly visible, yet it is significantly smaller and somewhat broader. In
contrast, in the C(T)  data of the Co-richest alloy UNi$_{0.02}$Co$_{0.98}$Si$_2$ hardly any
singularity is seen. The encountered evolution of the $\lambda$-peak is entirely in line with the
suppression of the ferromagnetic order in UNi$_{1-x}$Co$_x$Si$_2$ observed in the $\sigma (T)$ and
$\rho(T)$ characteristics.

Figure~\ref{fig_heat_all}(b) displays the temperature dependencies of the excess specific  heat
$\Delta C$ due to $5f$-electrons divided by $T$, obtained for a few UNi$_{1-x}$Co$_x$Si$_2$
subtracting the $C(T)$ data of nonmagnetic ThCoSi$_2$ from the heat capacity of these alloys. The
specific heat of the ferromagnetically ordered compounds can be described below about $0.8 \,
T_{\rm C}$ by the formula developed for thermally excited ferromagnetic magnons
(cf.~Ref.~\onlinecite{gopal}):
\begin{equation}
\Delta C(T) = \gamma^{\ast} T + B T^{3/2} \exp{\left( -\frac{\Delta}{T} \right)}.
\label{eq_hc_magnons}
\end{equation}
where $\gamma^{\ast}$ is the Sommerfeld coefficient due to partly delocalized
$5f$-electrons of uranium, $\Delta$ stands for the gap in the spin-wave spectrum, and $B$ is a
coefficient of proportionality. Least squares fits of Eq.~(\ref{eq_hc_magnons}) to the experimental
data yielded for UNiSi$_2$: $\gamma^{\ast}$~=~20(1)~mJ/(molK$^2$), $B$~=~36(1)~mJ/(molK$^4$) and
$\Delta$~=~81(1)~K, and for UNi$_{0.40}$Co$_{0.60}$Si$_2$: $\gamma^{\ast}$~=~42(1)~mJ/(molK$^2$),
$B$~=~50(1)~mJ/(molK$^4$) and $\Delta$~=~54(1)~K.

As can be inferred from Fig.~\ref{fig_heat_all}(b), in the $\Delta C(T)/T$ derived for
UNi$_{0.02}$Co$_{0.98}$Si$_2$  there is no $\lambda$-shaped anomaly, and only a broad hump occurs
in the vicinity of $T^{\ast}$~=~6~K, i.e. near the temperature of the anomaly evidenced in the
magnetization curve [compare Fig.~\ref{fig_suscept_all}(b)]. Consequently, this feature in the
specific heat data might be interpreted as a result of short range magnetic ordering or spin-glass
state formation.

Another interesting feature, well visible in the specific heat of UNi$_{0.02}$Co$_{0.98}$Si$_2$, is
a broad maximum located at about 30~K [Fig.~\ref{fig_heat_all}(b)]. Similar anomaly is present also
in pure UCoSi$_2$ [Fig.~\ref{fig_heat_all}(b)] and in Th-doped UCoSi$_2$. Preliminary results
obtained for the latter alloys (to be published elsewhere) show that the position of this anomaly
is independent of Th-content, yet its magnitude scales with the dilution of the uranium sublattice.
This finding clearly indicates a single-ion character of the hump, which probably results from
splitting the the $5f$ multiplet in crystalline-electric-field potential. Closer look at the
$\Delta(C)/T$ data allows to recognize some faint features at 30~K also in UNiSi$_2$ and
UNi$_{0.40}$Co$_{0.60}$Si$_2$ [Fig.~\ref{fig_heat_all}(b)], being in line with the latter
hypothesis. However, the Schottky contribution is here largely obscured by the magnetic
contribution due to the ferromagnetic ordering.

Figure~\ref{fig_heat_all}(c) displays the temperature dependencies of the increase in the magnetic
entropy $\Delta S$, defined as:
\begin{equation}
\Delta S = \int^T_{T_{\rm min}} \frac{\Delta C(T)}{T} {\rm d}T,
\label{eq_entropy}
\end{equation}
where $T_{\rm min}$ is the lower temperature limit in our experiments. As seen, $\Delta S$ in
UNi$_{1-x}$Co$_x$Si$_2$ increases with increasing temperature up to about 100~K and then saturates at
about 9~J/(mol K). The latter value is close to $R\ln 3$ (where $R$ is the universal gas constant),
which is expected for a thermally populated triplet or pseudotriplet. Assuming that the system
studied has a magnetic doublet as a ground state, the first excited level is a singlet, lying about
100~K above the ground level. Negligible increase of $\Delta S$ above 100~K suggests that the next
excited CEF level (of unknown degeneracy) lies well above 300~K. Similar magnitude of CEF splitting is often
observed in U-based systems.

The enlarged electronic contribution to the specific heat of UNiSi$_2$ and
UNi$_{0.40}$Co$_{0.60}$Si$_2$ is a fingerprint of strong electronic correlations. Based on the
value of the $\gamma^{\ast}$ coefficient both materials can be qualified as moderately-enhanced
Fermi liquids. Upon further increasing the Co-content the low-temperature heat capacity evolves
into $C/T \sim -\ln{T}$ dependence, especially well resolved in UCoSi$_2$, characteristic of
non-Fermi liquid systems.

\section{\label{summary} Summary}

\begin{figure}
\includegraphics[width=7cm]{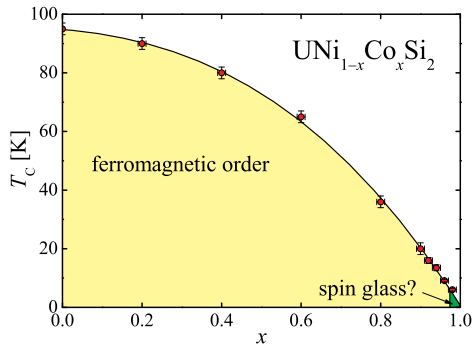}%
\caption{\label{fig_phase_diag} (Color online) Tentative magnetic phase diagram of UNi$_{1-x}$Co$_x$Si$_2$.
Solid line serves as a guide for the eye.}
\end{figure}

The presented experimental data revealed that the ferromagnetism and pronounced Kondo effect,
observed in UNiSi$_2$, is gradually suppressed upon stepwise substitution of Ni by Co. While in the
UNi$_{1-x}$Co$_x$Si$_2$ alloys with $x < 0.98$ the phase transition exhibits clearly the
ferromagnetic character, in the sample with $x = 0.98$ a spin-glass-like state with dominant spin
fluctuations is seen. Whether any short-range order occurs in UCoSi$_2$ below 2~K, remains an open
question.

Since the change of the unit cell volume of the system UNi$_{1-x}$Co$_x$Si$_2$ is relatively small
and has different sign along the main crystallographic axes, the partial Ni/Co substitution cannot
be treated as an approximation of hydrostatic pressure. Also the magnetic sublattice of uranium
ions remains unaltered in the presented experiments. Therefore, the observed evolution of the
physical properties of UNi$_{1-x}$Co$_x$Si$_2$ results most probably from electron doping, because
the electron configurations of nickel and cobalt differ from each other by one electron on the $3d$
shell.

Most interestingly, the extrapolation of the $T_{\rm C}(x)$ curve to $x \rightarrow 1$
(Fig.~\ref{fig_phase_diag}) suggests the stoichiometric UCoSi$_2$ compound to be near the
ferromagnetic quantum phase transition. Further experiments, performed at mK temperature range on
single crystalline samples of UCoSi$_2$, are needed to verify the latter hypothesis. In this
context,  particularly tempting seems a conjecture on the presence in UCoSi$_2$ of quantum critical
point governed by ferromagnetic fluctuations and exclusion of the possibility of a rapid change of
the character of the correlations in the alloys with $0.96 < x < 1$.

\begin{acknowledgments}
This work was supported by the Polish Ministry of Science and Higher Education within
grant no. N N202 102338.
\end{acknowledgments}

\end{document}